\documentclass[10pt, conference, letterpaper]{ IEEEtran}
\IEEEoverridecommandlockouts

\usepackage{amsmath,amsfonts, amsthm}
\usepackage{algorithmic}
\usepackage{graphicx}
\usepackage{textcomp}
\usepackage{xcolor}
\usepackage{acronym}
\usepackage{subfig}
\usepackage{balance}
\usepackage{booktabs}
\usepackage{tabularx}
\usepackage{hyperref}
\usepackage[left=0.6in,right=0.63in,top=0.75in, bottom=1in]{geometry}

\usepackage{tcolorbox}  

\usepackage[colorinlistoftodos,textwidth=\marginparwidth]{todonotes}

\newtheoremstyle{findingstyle}
    {0pt}                    
    {0pt}                    
    {\normalfont}            
    {0pt}                    
    {\bfseries}              
    {. }                     
    {0.0em}                  
    {}                       
\theoremstyle{findingstyle}

\newtheorem{findinginner}{Finding}[section]
\newenvironment{finding}
{\begin{tcolorbox}[
    colback=gray!10,              
    colframe=gray,                
    arc=5pt,                      
    boxrule=1pt,                  
    left=0pt,                     
    right=0pt,                    
    top=0pt,                      
    bottom=0pt,                   
    left skip=0pt,                
    right skip=0pt                
]%
\begin{findinginner}}
{\end{findinginner}%
\end{tcolorbox}}

\def\BibTeX{{\rm B\kern-.05em{\sc i\kern-.025em b}\kern-.08em
    T\kern-.1667em\lower.7ex\hbox{E}\kern-.125emX}}

\acrodef{PHY}{Physical}
\acrodef{RF}{Radio Frequency}
\acrodef{RFF}{Radio Frequency Fingerprinting}
\acrodef{RFFI}{Radio Frequency Fingerprint Identification}
\acrodef{ML}{Machine Learning}
\acrodef{DL}{Deep Learning}
\acrodef{BPSK}{Binary Phase Shift Keying}
\acrodef{SDR}{Software Defined Radio}
\acrodef{USRP}{Universal Software Radio Peripheral}
\acrodef{BPSK}{Binary Phase Shift Keying}
\acrodef{TED}{Time Error Detector}
\acrodef{FPGA}{Field Programmable Gate Array}
\acrodef{NN}{Neural Network}
\acrodef{CNN}{Convolutional Neural Networks}
\acrodef{AE}{Autoencoder}
\acrodef{CAE}{Convolutional Autoencoder}
\acrodef{DL}{Deep Learning}
\acrodef{IQ}{In-Phase Quadrature}
\acrodef{MSE}{Mean Squared Error}
\acrodef{LogMSE}{Logarithmic Mean Squared Error}
\acrodef{ROC}{Receiver Operating Characteristics}
\acrodef{AUC}{Area Under the Curve}
\acrodef{TPR}{True Positive Rate}
\acrodef{FPR}{False Positive Rate}
\acrodef{TNR}{True Negative Rate}
\acrodef{FNR}{False Negative Rate}
\acrodef{sps}{samples per second}
\acrodef{ID}{In-Distribution}
\acrodef{OOD}{Out-Of-Distribution}
\acrodef{LLM}{Large Language Model}
\acrodef{SDR}{Software Defined Radio}
\acrodef{MSE}{Mean Squared Error}
\acrodef{TAR}{True Acceptance Rate}
\acrodef{FAR}{False Acceptance Rate}
\acrodef{CFO}{Carrier Frequency Offset}
\acrodef{FFT}{Fast Fourier Transform}

\begin{document}

\title{Where You Tap Matters: A Probe-and-Model Benchmark for Open-Set RF Fingerprinting}

\author{
    \IEEEauthorblockN{Gabriele Oligeri\IEEEauthorrefmark{1}, Savio Sciancalepore\IEEEauthorrefmark{2}, Ingrid Huso\IEEEauthorrefmark{1}, Fatima Al-Mousawi~\IEEEauthorrefmark{1}}
    \IEEEauthorblockA{\IEEEauthorrefmark{2}Eindhoven University of Technology, Eindhoven, Netherlands.\\ s.sciancalepore@tue.nl}
    \IEEEauthorblockA{\IEEEauthorrefmark{1}College of Science and Engineering, Hamad Bin Khalifa University, Doha, Qatar.\\ \{goligeri, ihuso, faal30798\}@hbku.edu.qa}
}

\maketitle


\begin{abstract}
Radio Frequency Fingerprint Identification (RFFI) enables transmitter identification at the physical layer by learning device-specific impairments from received signals, yet the literature is inconsistent about where in the receiver chain those samples should be collected. Since distinct transformations are applied to the signal by the different receiver operations, i.e., carrier recovery, gain normalization, pulse shaping, and timing recovery, they can either tighten within-transmitter variability or suppress the features RFFI requires for classification. We present a systematic real-world evaluation of open-set, reconstruction-error RFFI using data collected at five probe points along a standard BPSK receiver chain. Our results show that RFFI is strongly probe-dependent: timing recovery and, to a lesser extent, carrier recovery enable low false-acceptance operation with limited in-distribution–out-of-distribution overlap, whereas other stages often require a false-acceptance ratio above 0.1 to achieve a true-acceptance ratio of 0.9. To test the validity of our findings across model selection, we benchmark several LLM-designed autoencoders using a controlled pipeline that holds preprocessing and MSE scoring fixed. These architectures confirm that RFFI is probe-dependent. Moreover, they do not outperform the baseline at the chosen operating point and typically increase training time. Overall, probe selection dominates reconstruction-based open-set RFFI performance, more than the autoencoder complexity.
\end{abstract}


\section{Introduction}
\ac{RFFI} addresses the problem of transmitter identification by exploiting device-dependent characteristics embedded in received baseband IQ samples~\cite{jagannath2022_comnet}. Rather than relying on higher-layer identifiers, \ac{RFFI} aims to infer which device generated a radio signal, enabling device-level identification even when protocol metadata, e.g., packet headers, are unavailable or untrusted. In security-relevant settings, this problem naturally maps to detecting whether an incoming waveform, typically collected through a \ac{SDR}, has a \emph{profile} consistent with an enrolled device or an unauthorized entity. Thus, \ac{RFFI} can be formulated as an open-set decision problem driven by how well a learned model reconstructs ``known'' signals~\cite{smailes2025satiq}. Many scientific contributions in recent years have focused on \ac{RFFI}, in the context of several radio communication technologies~\cite{xu2016_comst}. The largest corpus of the scientific literature focuses on the design of advanced signal processing techniques to enhance \ac{RFFI} performance. Pioneer approaches used pure statistical methods~\cite{bihl2016_tifs}. More recently, the domain received significantly more attention from the community with the increasing diffusion and popularity of \ac{DL} algorithms, particularly \ac{NN} classifiers~\cite{saeif_day-after-tomorrow_2023} and anomaly-based approaches like \acp{AE}~\cite{wigchert2026_comnet}. At the same time, each proposed \ac{RFFI} solution applies such techniques on \ac{PHY} data collected at various stages of the receiver chain, e.g., immediately after the antenna~\cite{smailes_watch_2023}, after the \ac{FFT}~\cite{al2020exposing}, or after timing recovery~\cite{oligeri_past-ai_2023}. Although the literature provides some studies on classifier-driven performance evaluation for \ac{RFFI}, the impact of the data collection stage has received no attention. Several signal processing operations are executed on a received radio signal, e.g., carrier recovery, gain normalization, pulse shaping, and timing recovery. These operations reshape the structure of the \ac{IQ} data used for \ac{RFFI}: they could reduce within-transmitter variability (helping separability) but could also suppress or homogenize device-specific features, negatively affecting \ac{RFF}. Overall, the ``best'' input data for \ac{RFFI} is not guaranteed to be the earliest raw samples, i.e., the immediate output taken from the radio interface, nor the output of the receiver chain. Consequently, the \ac{DL}-based algorithm used for \ac{RFFI} should be evaluated in the context of the probe at which it operates.

{\bf Contribution.} In this work, we present a systematic experimental analysis of the impact of the data collection probe on the performance of state-of-the-art open-set \ac{RFFI}. We collect \ac{IQ} samples emitted from twelve (12) \ac{SDR} transmitters at five probes along a representative GNU Radio receiver chain, each implementing a communication-oriented transformation, i.e., carrier recovery, gain normalization, pulse shaping, and timing recovery. Using state-of-the-art image-based \acl{AE}-driven \ac{RFFI} algorithms, we demonstrate quantitatively that \ac{RFFI} performances change significantly based on the data collection stage. To remove potential biases due to the configuration of the chosen classifier, we also use publicly-available \acp{LLM} to generate several benchmark configurations to be compared with the current state-of-the-art baseline. The LLM-generated designs span different directions: stronger sparsity and smaller bottlenecks (GPT5.2), denoising plus sparsity (Gemini 3.1 Pro), deeper convolutional \acp{AE} with normalization/dropout and learning-rate scheduling (GLM5), compact convolutional \acp{AE} (SuperGrok), and convolutional \acp{AE} with custom training and an explicit latent-structure regularizer (Claude Sonnet 4.6). All such LLM-generated networks report similar performance trends to the baseline, demonstrating the robustness of our findings to the configurations of the chosen \ac{RFFI} technique. Moreover, the generated networks do not improve \ac{RFFI} performances over the baseline, while introducing additional training overhead. 

In summary, our paper provides the following contributions:
\begin{itemize}
    \item We deliver a systematic, probe-aware evaluation of image-based open-set \ac{RFFI} across five receiver-chain observation points and twelve transmitters, revealing that reliable fingerprinting is strongly probe-dependent.
    \item We show that probe choice can affect the resulting \ac{ID}/\ac{OOD} reconstruction-error separation: timing recovery and carrier recovery provide substantially better operating points than earlier or intermediate stages in our controlled \ac{BPSK} receiver chain.
    \item To check whether the probe-dependent trend is an artifact of one particular \ac{AE} implementation, we additionally compare the baseline against a small set of alternative \ac{AE} designs generated by contemporary \ac{LLM} tools. We do not claim to evaluate \ac{LLM}-assisted neural architecture search in general; rather, these designs serve as an architectural-diversity stress test under a fixed preprocessing and scoring pipeline. 
    \item As a secondary robustness check, we compare the baseline \ac{AE} with several alternative \ac{AE} designs produced by the \acp{LLM} while holding preprocessing and \ac{MSE} scoring fixed. These alternatives do not change the main probe-dependent conclusion and generally increase the training cost.
\end{itemize}

Overall, our work demonstrates that the achievement of strong \ac{RFFI} discrimination in practice depends critically on two coupled design choices: (i) \emph{where} in the receiver chain the signal is observed, and (ii) \emph{how} the learned model is constructed. Probe selection dominates performance, while most \ac{LLM}-driven architectural changes do not remove probe dependence, while providing limited performance benefit at the chosen operating point. We also clarify the practical trade-off that more complex designs increase training costs with little (and often negative) impact on discrimination.

{\bf Roadmap.} The rest of this paper is organized as follows. Section~\ref{sec:related_work} reviews prior work on \ac{RFFI}, Section~\ref{sec:data_collection} describes our data collection setup, Section~\ref{sec:transmitter_identification} formalizes the open-set \ac{RFFI} problem, Section~\ref{sec:methodology} defines the evaluation methodology, Section~\ref{sec:LLMassistedNN} introduces our considered \ac{AE} designs, Section~\ref{sec:performance_evaluation} reports performance and timing results, Section~\ref{sec:probes} studies probe-aware enrollment and cross-probe transferability, Section~\ref{sec:discussion} discusses implications, trade-offs, and limitations, and finally, Section~\ref{sec:conclusion} concludes the paper.

\section{Related work}
\label{sec:related_work}

\acl{RFFI} has historically investigated transmitter identification~\cite{irfan_radio_2025,al-hazbi_radio_2024} by exploiting hardware-dependent impairments observable in the received signal, including effects such as \ac{IQ} imbalance~\cite{aqib_khan_comprehensive_2025, yao_novel_2024, niu_identification_2023}, phase shift~\cite{yang_phase-proof_2025}, and \ac{CFO}~\cite{wheeler_assessment_2017}, often in combination with conventional \ac{ML} models. With the diffusion of \ac{DL}, many \ac{RFFI} approaches shifted toward learning directly from raw physical-layer observations, i.e., baseband \ac{IQ} samples, leveraging the ability of \acp{NN} to extract hidden patterns and achieve very strong classification performance~\cite{al-hazbi_radio_2024}. \ac{DL} models used in this context commonly include \ac{CNN}-based classifiers~\cite{yan_radio_2025} and \ac{AE}-based methods~\cite{yu_radio_2019} operating on \ac{IQ}-derived representations such as images.

A recurring finding across recent literature is that high-performing deep \ac{RFFI} models can be unstable under real-world variability~\cite{saeif_day-after-tomorrow_2023}. Their outputs may be significantly affected by noise and non-stationarity introduced by practical operating conditions such as radio warm-up~\cite{elmaghbub_no_2024}, hardware resets~\cite{saeif_day-after-tomorrow_2023}, temperature changes~\cite{nieves_measurements_2026}, and time-varying channel conditions~\cite{cai_rff_2025}, among others. This sensitivity has motivated a substantial body of work targeting robustness across deployment conditions, including explicit handling of distortions and the evaluation of performance over time.

One prominent approach is constituted by image-based \ac{RFFI}, in which raw \ac{IQ} samples are converted into image-like representations before learning. Prior work shows that preprocessing \ac{IQ} into images can yield models that are more robust across disruptive factors and operational changes, e.g., adversarial perturbations~\cite{papangelo_adversarial_2024}, radio restarts~\cite{saeif_day-after-tomorrow_2023}, firmware reloads~\cite{irfan_shifting_2025}, and similar image-based techniques have also been applied beyond device identification, including channel fingerprinting~\cite{oligeri_satprint_2024,oligeri_fadeprint_2024} and jamming detection~\cite{alhazbi_bloodhound_2023}. 

Although many prior \ac{RFFI} works treat transmitter identification as a closed-set classification task~\cite{oligeri_past-ai_2023}, security-relevant scenarios naturally motivate open-set decisions: determining whether an incoming signal is consistent with an enrolled transmitter rather than an unauthorized one~\cite{huang_radio_2024}. A common solution is reconstruction-based identification using an \ac{AE}: train an \ac{AE} on \ac{ID} samples, then use reconstruction error (MSE) as a scalar score to separate \ac{ID} from \ac{OOD} samples. Importantly, reconstruction-based \ac{RFF} introduces a key design issue: architectural changes that improve reconstruction quality ``in general'' can also improve reconstruction of \ac{OOD} inputs, potentially shrinking the \ac{ID}–\ac{OOD} error gap on which open-set discrimination relies~\cite{Zhou_2022_CVPR,xiao_likelihood_2020}.

Another practical aspect, often under-emphasized, is that the received signal undergoes significant reshaping during the receiver's processing stages. Thus, the stage in the reception process where data is collected is important. In most papers, the data collection sampling point is not reported at all. In other cases, the literature is not aligned. For example, the work in~\cite{smailes_watch_2023} collected IQ samples immediately after baseband conversion, while the authors in~\cite{irfan_shifting_2025} collected them after timing recovery. 
Nonetheless, to our knowledge, no prior work has specifically investigated the extent to which the \ac{IQ} collection probe can affect \ac{RFFI} performance.

Given the size and coupling of this design space, i.e., the combination of probes and transmitters, recent works~\cite{tayebi_arasteh_large_2024,nasir} suggest that \acp{LLM} can serve as code- and design-generation assistants to improve inference performance. To the best of our knowledge, no prior literature has investigated the extent to which mainstream available \acp{LLM} can improve the design and code of \ac{RFFI} techniques. We address both mentioned gaps throughout this paper.

\section{Data Collection}
\label{sec:data_collection}

In this section, we introduce the measurement methodology used to build the \ac{RF} dataset for our work. 

\begin{figure*}
 \centering
 \includegraphics[width=\textwidth, angle = 0,trim = 0mm 50mm 0mm 50mm]{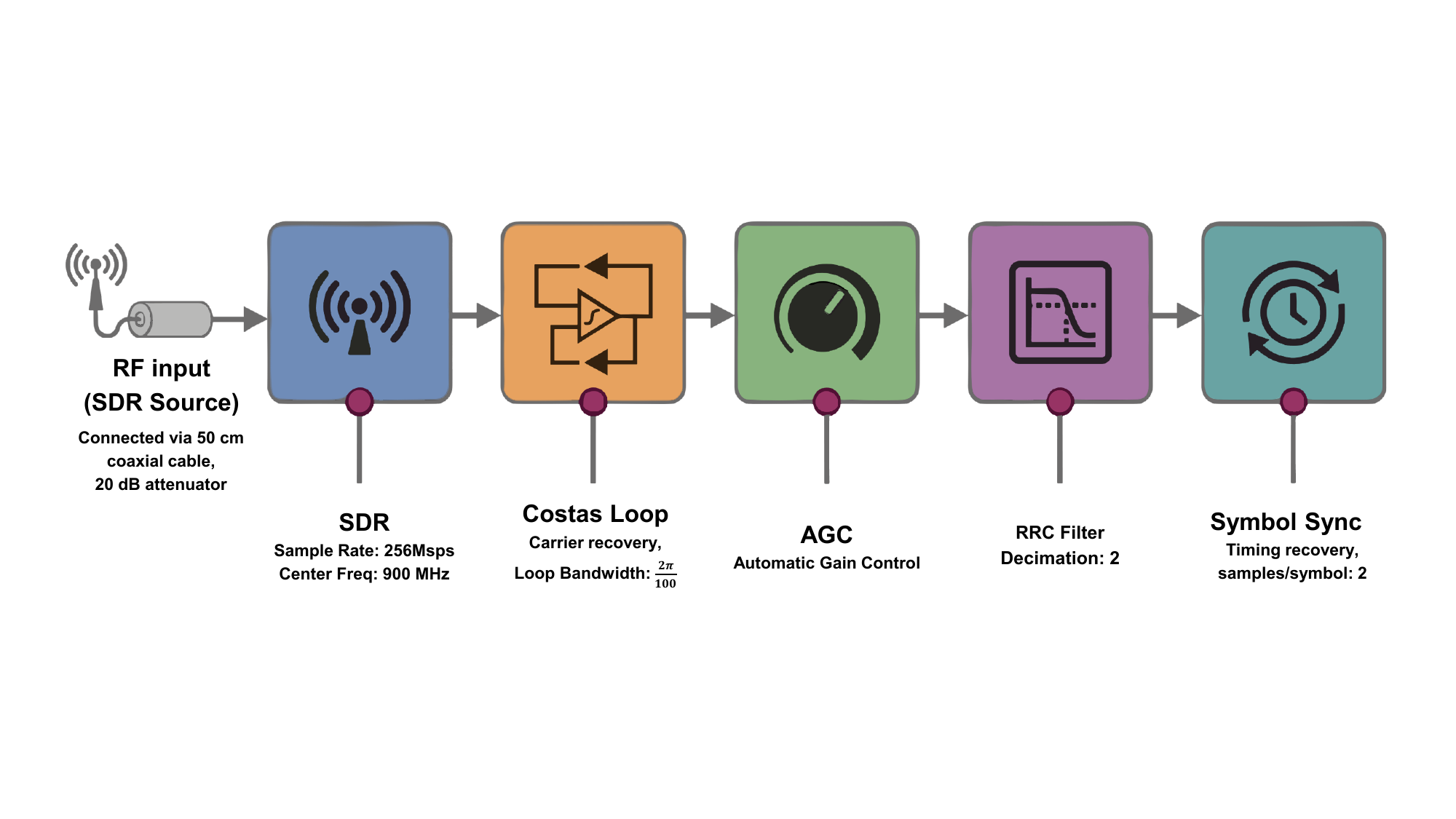}
 \caption{Schematic representation of the GnuRadio receiver chain and points (probes) where the IQ samples are taken: \texttt{radio}, \texttt{costasloop}, \texttt{agc}, \texttt{rrc}, and \texttt{symbsync}.}
 \label{fig:receiver_chain}
\end{figure*}

{\bf Hardware.} Our measurement setup consists of thirteen (13) \acp{SDR}: twelve USRP B200-mini-i devices used as transmitters and one USRP X410 used as the receiver. Each transmitter is connected to the receiver through a 50~cm RF coaxial cable and a 20~dB attenuator. All experiments are performed over cable rather than over the air to eliminate wireless channel effects, such as multipath and fading, that could otherwise influence our observations. This controlled and isolated setup allows us to focus exclusively on device-level phenomena, without interference from wireless channel artifacts.

{\bf Software.} All radios are controlled through GNU Radio 3.8, while \ac{RFF} transmitter identification is performed in MATLAB R2023b. We implemented a \ac{BPSK} standard transmitter/receiver chain without relying on higher-layer protocols. This design choice avoids MAC-layer orchestration effects, such as silences and bit-pattern regularities, thereby maximizing the observability of RFF-related characteristics. At the transmitter, a GNU Radio Vector Source continuously emits a deterministic byte sequence, \{0,…,255\}, which is mapped into complex \ac{BPSK} symbols at a rate of 128,000 samples per second (sps). The transmit chain then applies a Root Raised Cosine (RRC) filter with interpolation factor 4, gain 2, roll-off factor 0.35, and 128 taps, before feeding a USRP Sink configured with a sampling rate of 256,000 sps, normalized transmit power of 0.5, and center frequency of 900 MHz. At the receiver, a USRP Source is configured at 900 MHz and 256,000 sps with a normalized gain of 0.5. The received signal is then processed through a Costas Loop and an Automatic Gain Control (AGC) block, followed by a matched RRC filter with the same parameters as the transmitter-side filter and a decimation factor of 2. Finally, symbol timing recovery is performed through a Symbol Sync block using a Mueller and Müller Time Error Detector (TED) operating at 2 sps. The normalized transmitter power (0.5), receiver gain (0.5), and 20 dB attenuation were selected empirically to ensure long-term measurement stability. To inspect the signal evolution across the receiver chain, we placed multiple \emph{File Sink} blocks, hereafter referred to as \textit{probes}, at different processing stages, as shown in Fig.~\ref{fig:receiver_chain}.  Specifically, a File Sink is placed at the output of each processing block. In the following, the resulting IQ datasets are referred to as \texttt{radio}, \texttt{costasloop}, \texttt{agc}, \texttt{rrc}, and \texttt{symbsync}, respectively.

{\bf Dataset.} Table~\ref{tab:datasets} summarizes the datasets collected for our work. For each transmitter, we acquired 5 measurements, each lasting 10 minutes. This results in a total of approximately $256,000 \times 60 \times 10 \approx 153$~million IQ samples for the \texttt{radio}, \texttt{costasloop}, and \texttt{agc} datasets, where no decimation is applied. For the \texttt{rrc} dataset, a decimation factor of 2 reduces the total to 76 million IQ samples. A further decimation factor of 2 is then applied within the \texttt{symbsync} block to produce the \texttt{symbsync} dataset, yielding 38 million IQ samples.
\begin{table}
\centering
\caption{Summary of our Datasets.}
\begin{tabular}{l|c|c}
\textbf{\begin{tabular}[c]{@{}c@{}}Probe \end{tabular}} & \textbf{\begin{tabular}[c]{@{}c@{}}IQ \\ samples [M]\end{tabular}} & \textbf{\begin{tabular}[c]{@{}c@{}}Decimation\\ factor\end{tabular}} \\ \hline
\textit{radio}      & 153 & \multicolumn{1}{c}{1} \\ \hline
\textit{costasloop} & 153 & \multicolumn{1}{c}{1} \\ \hline
\textit{agc}        & 153 & \multicolumn{1}{c}{1} \\ \hline
\textit{rrc}        & 76  & \multicolumn{1}{c}{2} \\ \hline
\textit{symbolsync} & 38  & \multicolumn{1}{c}{2} \\ \hline
\end{tabular}
\label{tab:datasets}
\end{table}
Finally, we verified that the bit error rate at the output of the \texttt{symbsync} block is zero, confirming that all transmitted bits were received correctly. The dataset and the scripts adopted in this paper will be made publicly available upon acceptance.

\section{Transmitter Identification}
\label{sec:transmitter_identification}

We formulate the transmitter identification task as an \emph{open-set classification} problem.  Let us define as $D$ one of the datasets described in Sec.~\ref{sec:data_collection}, i.e., the output of the five probes placed along the receiver chain, for a total of 60 measurements. In the remainder of this paper, we follow the standard \ac{AE} training-reconstruction procedure. The reference dataset $D$ is divided into two subsets, i.e., $D = \{D_a, D_b\}$, where $D_a$ corresponds to 70\% of the dataset and is used to train the model $\alpha$, namely $\alpha = T(D_a)$. The trained model $\alpha$ is then used to reconstruct the remaining portion of the dataset, $D_b$, producing $D_b'$. We highlight that $D_b$ might come from the same transmitter, i.e., \ac{ID}, or from another one, i.e., \ac{OOD}. Finally, we compute the associated \ac{MSE} between the original and reconstructed datasets, i.e., between $D_b$ and $D_b'$. 

\begin{figure*}
    \centering
    \begin{minipage}{0.19\textwidth} 
        \centering
        \includegraphics[width=\linewidth,trim = 10mm 60mm 25mm 60mm]{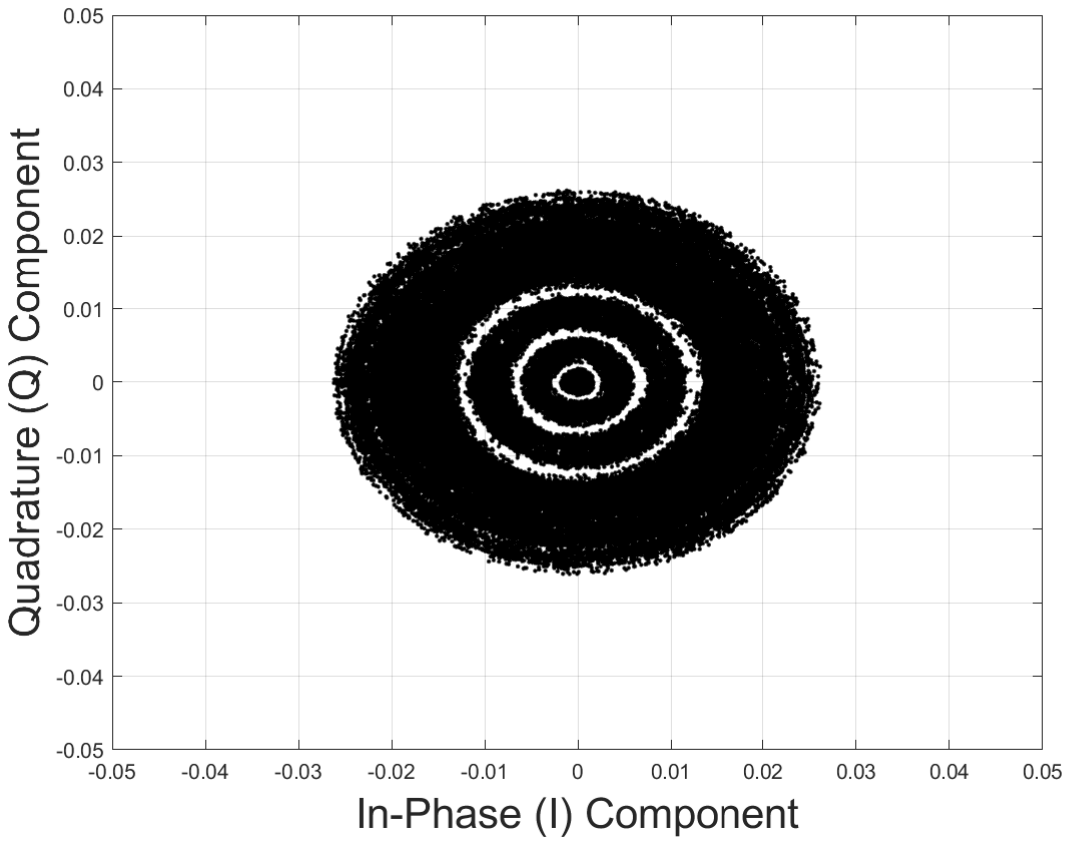}
        \caption*{(a)}
        \label{fig:figure1}
    \end{minipage}%
    \hfill 
    \begin{minipage}{0.19\textwidth} 
        \centering
        \includegraphics[width=\linewidth,trim = 10mm 60mm 25mm 60mm]{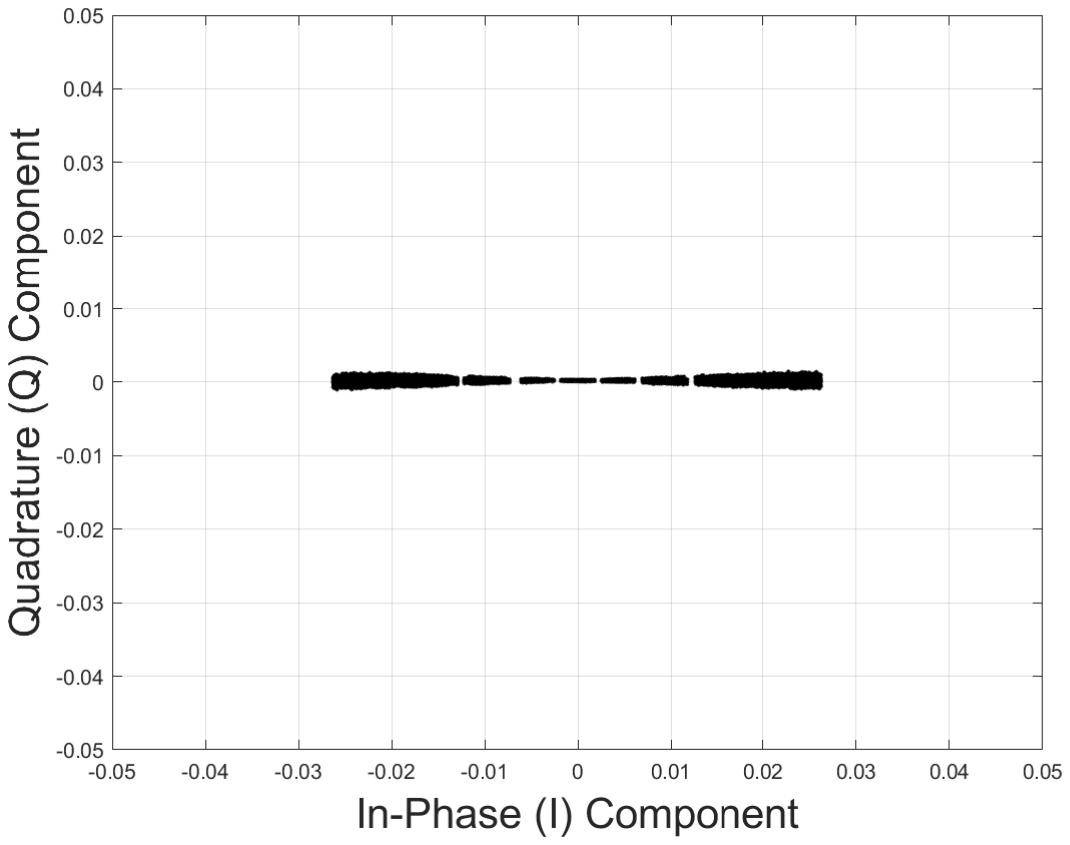}
        \caption*{(b)}
        \label{fig:figure2}
    \end{minipage}
    \hfill 
    \begin{minipage}{0.19\textwidth} 
        \centering
        \includegraphics[width=\linewidth,trim = 10mm 60mm 25mm 60mm]{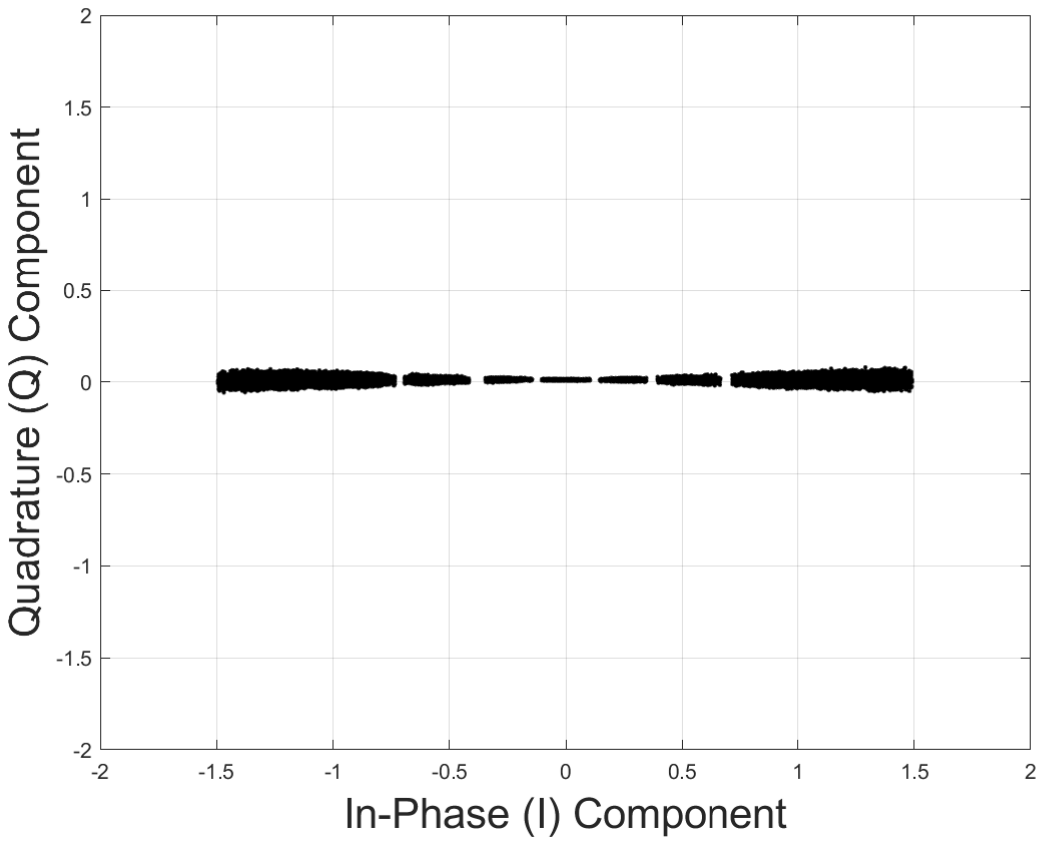}
        \caption*{(c)}
        \label{fig:figure3}
    \end{minipage}
    \hfill 
    \begin{minipage}{0.19\textwidth} 
        \centering
        \includegraphics[width=\linewidth,trim = 10mm 60mm 25mm 60mm]{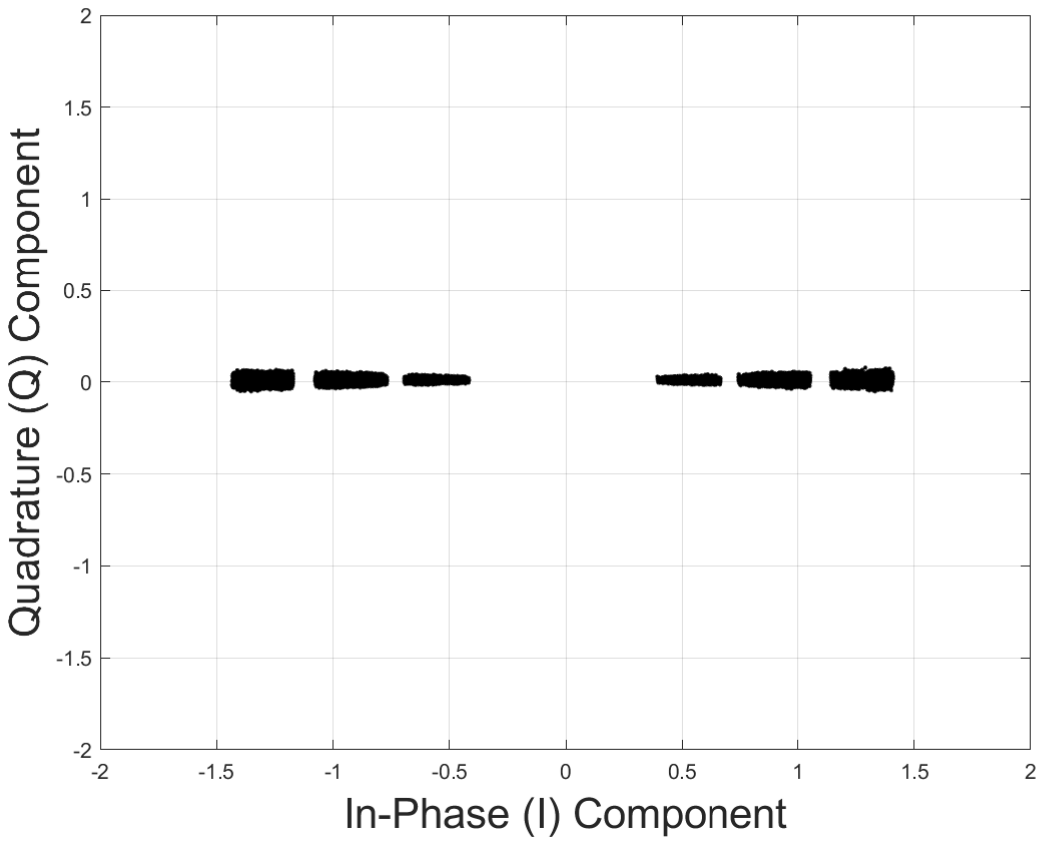}
        \caption*{(d)}
        \label{fig:figure4}
    \end{minipage}
    \hfill 
    \begin{minipage}{0.19\textwidth} 
        \centering
        \includegraphics[width=\linewidth,trim = 10mm 60mm 25mm 60mm]{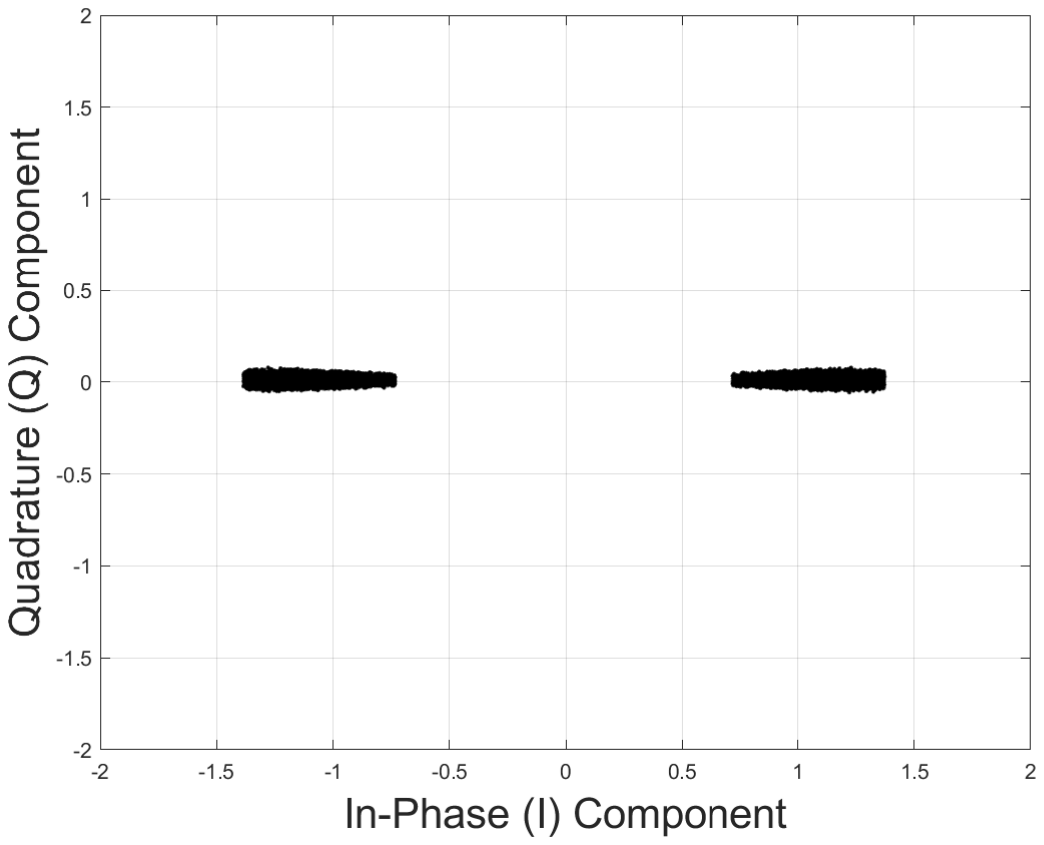}
        \caption*{(e)}
        \label{fig:figure5}
    \end{minipage}    
    \caption{IQ samples visualization example. We report $10^5$ IQ samples taken from the probes \texttt{radio} (a), \texttt{costasloop} (b), \texttt{agc} (c), \texttt{rrc} (d), and finally, \texttt{symbsync} (e).}
    \label{fig:iq_examples}
\end{figure*}

Figure~\ref{fig:iq_examples} shows an example of $1 \cdot 10^5$ \ac{IQ} samples collected from each of the probes along the receiver chain of Fig.~\ref{fig:receiver_chain}. 
At the \verb|radio| probe, the samples are distributed in a circular region around the origin, reflecting the combined effects of modulation, the RF cable channel, hardware impairments, and noise. After carrier recovery, i.e., at the \verb|costasloop| probe, the samples are rotated so that most of the energy lies on the in-phase component $I$, resulting in a thin horizontal line. This operation can be expressed as $x_{\mathrm{costas}}(n) = x_{\mathrm{radio}}(n)\,e^{-j\phi(n)}$, where $\phi(n)$ is iteratively updated to minimize the phase error. Through such a strategy, the loop compensates for carrier phase and frequency offsets by suppressing the $Q$ component, thereby collapsing the original circular structure. At the \texttt{agc} probe, the signal power is normalized over time, which reduces amplitude variability, i.e., fluctuations on the $I$ component, and consequently removes gain-related differences. The RRC filter then processes the signal (generating the signal available at the \texttt{rrc} probe), imposing an ideal pulse shape by linearly filtering the output of the \texttt{agc} probe, smoothing high-frequency and transient components and further regularizing the IQ distribution. Finally, at the \texttt{symbsync} probe, the symbol synchronizer samples the symbol centers, projecting the continuous-time waveform onto a set of points that again cluster along a narrow horizontal line, i.e., $\{-1,1\}$ according to the adopted \ac{BPSK} modulation. Overall, Fig.~\ref{fig:iq_examples} shows that each GNU Radio block in the receiver chain (Fig.~\ref{fig:receiver_chain}) fulfills a dedicated communication-oriented objective, namely, \emph{phase alignment}, \emph{power normalization}, \emph{pulse shaping}, and \emph{timing recovery}. In doing so, each block transforms the signal from the previous probe. As a consequence, the IQ structure (and the fingerprint of the transmitter) observed at one stage is not preserved, but is instead reshaped or partially removed by the subsequent processing stages.

{\bf Baseline configuration.} The objective of this paper is not to maximize absolute \ac{RFF} accuracy, but to isolate the effect of probe location under a fixed and reproducible reconstruction-based pipeline. Accordingly, we consider a state-of-the-art image-based \ac{AE} ({\em baseline}) used in several recent \ac{RFFI} works, e.g.,~\cite{oligeri_past-ai_2023,saeif_day-after-tomorrow_2023,irfan_shifting_2025,oligeri2026_wisec}. The baseline follows an image-based \ac{RFFI} pattern: it converts \ac{IQ} samples into compact 2D histogram-like representations, trains an unsupervised \ac{AE} on enrolled-device samples, and uses reconstruction error for open-set scoring. We therefore treat the baseline as a controlled reference implementation rather than as a newly proposed \ac{RFF} architecture. This choice keeps the probe comparison interpretable: changes in performance can be attributed primarily to the probe and not to probe-specific model tuning. Therefore, we adopt a standard pre-processing pipeline aligned with state-of-the-art image-based \ac{RFFI}. The baseline configuration implements an unsupervised, per-(TX, probe) shallow autoencoder trained on $32 \times 32$ \ac{RF} ``images'' derived from IQ data (as discussed in Sect.~\ref{sec:transmitter_identification}). Each image is produced by slicing the IQ stream into chunks of $100,000$ samples, mapping the pair $(|I|, Q)$ into a 2‑D histogram (\verb|histcounts2|) using (consistent) bin edges from the $5^{th}$–$95^{th}$ percentiles of the training data, applying \verb|log(1 + x)| compression, and normalizing to an 8‑bit range, i.e., $[0, 255]$. We highlight the importance of considering the absolute value of the $I$ component. As an example, considering the \verb|symbsync| probe, Fig.~\ref{fig:iq_examples}(e), the \ac{BPSK} constellation concentrates into two tight clusters around ($I \approx - 1$) and ($I \approx +1$) with ($Q \approx 0$). If we build the 2‑D histogram on the whole I-Q plane directly, these two clouds fall at opposite sides of the $I$-axis range. With a coarse $32 \times 32$ binning, most bins between the clouds are never populated, producing an almost empty image (all zeros) with only two small ``activated'' regions. Such sparse, low-occupancy inputs are poorly informative for the \ac{AE}.

During training, all images are loaded, vectorized to a $32 \times 32$ grid ($1024$ pixels), scaled to $[0,1]$, and concatenated as columns to form the final classifier input. The network is trained with the standard \verb|trainAutoencoder| function using a 64‑unit bottleneck, 300 epochs, L2 weight regularization of 0.001, sparsity regularization of 0.005, and the \verb|logsig| transfer functions in both the encoder and the decoder. At test time, the same image pipeline (with the same bin edges computed during training) is used; the model reconstructs using the \verb|predict| function, and the per-image reconstruction error is computed using the \ac{MSE}. 

\section{Methodology}
\label{sec:methodology}

Our main evaluation is ``single-enrolled'' and probe-fixed: one reference entity $(\bar{T}, \bar{P})$ is enrolled, with $\bar{T} \in \{1, \ldots, 12\}$ representing the transmitter, and $\bar{P} \in \{\verb|radio|, \verb|costasloop|, \verb|agc|, \verb|rrc|, \verb|symbsync|\}$ representing the probe.

For each enrolled pair $(\bar{T}, \bar{P})$, the model is trained only on training samples from transmitter $\bar{T}$ collected at probe $\bar{P}$. At test time, \ac{ID} samples are held-out samples from the same pair $(\bar{T}, \bar{P})$, while \ac{OOD} samples are samples from all other transmitters collected at the same probe, i.e., $(T, \bar{P})$, with $T \neq \bar{T}$. This strategy isolates transmitter discrimination at a fixed receiver-chain observation point. Cross-probe generalization is studied separately in Sect.~\ref{sec:probes}.

\begin{figure}
 \centering
 \includegraphics[width=0.95\columnwidth, angle = 0,trim = 0mm 0mm 0mm 0mm]{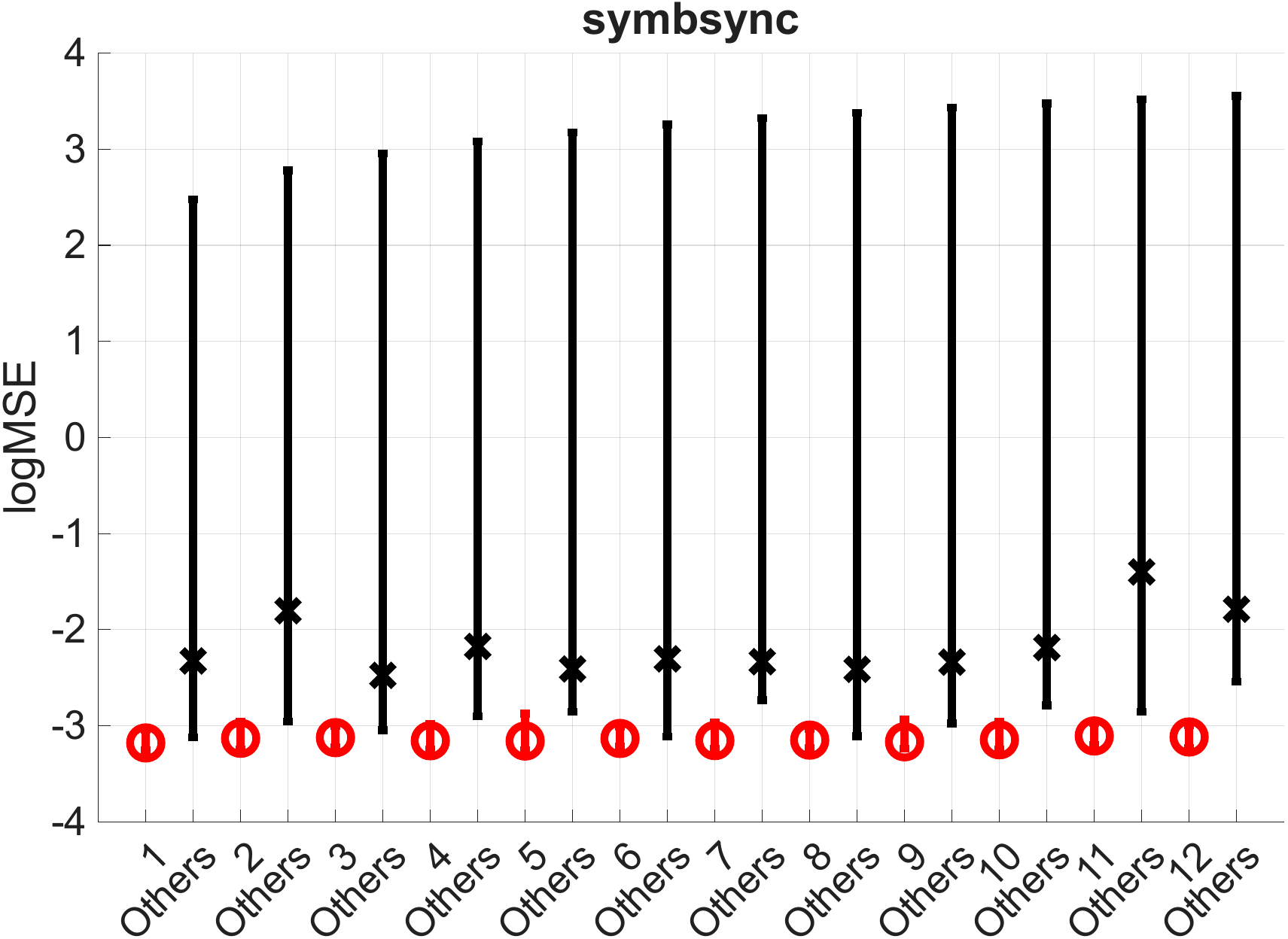}
 \caption{$\bar{\epsilon}$ as a function of all the possible pairs (Transmitter ID, Others) when considering the baseline design. Error-bars show the minimum, maximum, and median of the $\bar{\epsilon}$ while considering In-distribution (red) and Out-of-Distribution (black) samples.}
 \label{fig:symbsync}
\end{figure}

Each chunk of IQ samples ($10^5$) is mapped to a 2D image representation $x$ and fed into the autoencoder $\mathrm{AE}(\circ)$ to produce a reconstruction $\hat{x}$. We use the reconstruction \ac{MSE} as a scalar score, i.e., $\bar{\epsilon} = \log(\mathrm{MSE}(x,\hat{x}))$, where $\hat{x}=\mathrm{AE}(x)$. Figure~\ref{fig:symbsync} shows $\bar{\epsilon}$ as a function of all the possible comparisons between a reference transmitter $T \in \{1, \ldots, 12\}$ and the remainder of the pool ({\em Others}), when considering only the \verb|symbsync| probe. The error bars show the minimum, maximum, and median associated with the reconstructions $\bar{\epsilon}$, while considering the red and black bars for the \ac{ID} and \ac{OOD} samples, respectively. We note that there is only minor overlap between the red and black error-bars. Thus, the \verb|symbsync| probe is a good candidate for \ac{RFFI} when considering the baseline methodology.
In the following, we investigate the performance in more detail using the \ac{ROC} analysis. To this aim, we consider two metrics: \ac{TAR} and \ac{FAR}. The \ac{TAR}, or \ac{TPR}, measures the classifier's performance to correctly identify the reference entity, while the \ac{FAR}, or \ac{FPR}, measures the classifier's performance to wrongly accept the identity of an unauthorized entity.

We stress that we use the term open-set in the standard operational sense that the model is trained only on samples from an enrolled transmitter-probe pair and must reject samples from non-enrolled transmitters. This methodology should not be interpreted as an exhaustive open-world deployment with arbitrary numbers of unseen devices, signals, receiver implementations, or channels. Indeed, our dataset contains twelve SDR transmitters and is intended to isolate the effect of receiver-chain probe location under controlled conditions. Therefore, the \ac{FAR}/
\ac{TAR} values reported below should be read solely as comparative probe-dependent evidence within this measurement campaign, not as a universal deployment-scale false-acceptance estimate.

\begin{figure}
 \centering
 \includegraphics[width=0.95\columnwidth, angle = 0,trim = 0mm 0mm 0mm 0mm]{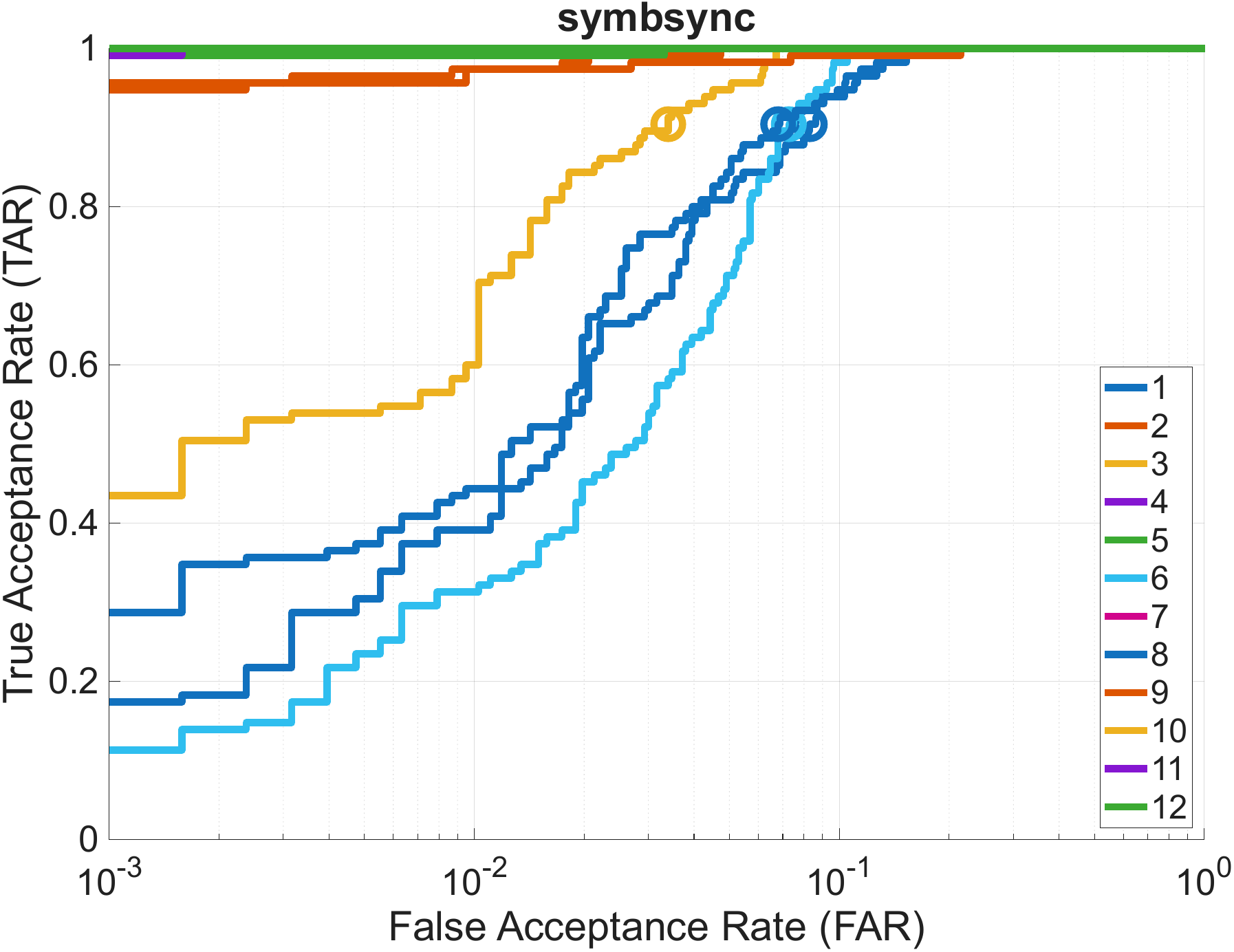}
 \caption{$\textrm{TAR}$ as a function of the $\textrm{FAR}$ while considering the {\em baseline} neural network design.}
 \label{fig:roc}
\end{figure}

We characterize the one-vs-others classification performance by sweeping the decision threshold $\tau$ applied to the $\bar{\epsilon}$. To this aim, we use a {\em lower-is-better} criterion, i.e., we ensure that lower reconstruction errors are mapped to higher scores. For each $\tau$, we compute the $\textrm{TAR}(\tau)$ and $\textrm{FAR}(\tau)$, then plot  $\textrm{TAR}$ (y-axis) versus $\textrm{FAR}$ (x-axis). Figure~\ref{fig:roc} shows the results for the \verb|symbsync| probe. The vast majority of transmitters can be identified ($\textrm{TAR} \approx 1$) with $\textrm{FAR} = 0$. Conversely, only 4 (out of 12) transmitters feature $\textrm{TAR} = 0.9$ while requiring $\textrm{FAR} < 0.09$. In the remainder of our analysis, we set $\textrm{TAR} = 0.9$ as a quality reference for the classifier and compare the $\textrm{FAR}$ required to achieve such a value as a criterion for feasible \ac{RFFI}.

We stress that the adoption of $\textrm{FAR}$ at $\textrm{TAR} = 0.9$ should be considered as an operating-point summary rather than an universal deployment target. Operationally, $\textrm{TAR} = 0.9$ means that the system accepts 90\% of held-out samples from the enrolled transmitter; the reported $\textrm{FAR}$ is then the fraction of non-enrolled transmitter samples that would be incorrectly accepted at that same threshold. Lower values of $\textrm{FAR}$ at this set value of $\textrm{TAR}$ indicate a cleaner separation between enrolled and non-enrolled reconstruction errors. We use this metric primarily for relative comparisons across probes and \ac{AE} designs. For deployment purposes, specific values of $\textrm{FAR}$ can be set according to operational requirements.

\section{LLM-Assisted Neural Network Design}
\label{sec:LLMassistedNN}
We consider several options to decorrelate our findings and conclusions from the specific configuration of the image-based \ac{AE} ({\em baseline}). To this aim, we investigate the extent to which various publicly available \acp{LLM} could design enhanced neural networks suitable for open-set \ac{RFFI}. We provide a reference implementation, i.e., our {\em baseline}, to five selected \acp{LLM}, representing diverse commercial players in the \ac{LLM} ecosystem: (i) {\em GPT5.2}, (ii) {\em Claude Sonnet 4.6}, (iii) {\em GLM5}, (iv) {\em Gemini 3.1 Pro}, and (v) {\em SuperGrok}. We considered the following prompt: {\em ``The enclosed scripts represent a baseline to perform radio frequency fingerprinting on MATLAB 2023b with all the toolboxes installed. Improve the performance of the classifier in terms of discrimination (accuracy). You can arbitrarily change the implementation of \verb|LLMTrain(obj, trainSet)| by considering (if needed) any type of network structure. The other parts of the code should stay the same."} 
We arranged the code in order to isolate the definition of the \ac{AE} structure inside the function \verb|LLMTrain(|$\circ$\verb|)|, thus enforcing the same data pre-processing and post-processing for all the considered \acp{AE}. We stress that our main objective is to investigate the feasibility of using \ac{LLM} to generate \acp{AE} that maximize the accuracy of the \ac{RFFI}.

{\bf LLM-assisted design of \ac{AE}.} Table~\ref{tab:llm_comparison} summarizes the main differences between the suggested LLM-based designs and our provided baseline.
\begin{table*}[t]
\centering
\caption{Comparison of our \ac{RFFI} baseline with other LLM-driven designs (\ac{DL} component).}
\label{tab:dl_compare}
\footnotesize
\setlength{\tabcolsep}{3.5pt}
\renewcommand{\arraystretch}{1.15}
\begin{tabularx}{\textwidth}{|>{\raggedright\arraybackslash}p{2.2cm}|
                             >{\raggedright\arraybackslash}p{3.6cm}|
                             >{\raggedright\arraybackslash}p{2.6cm}|
                             >{\raggedright\arraybackslash}X|
                             >{\raggedright\arraybackslash}p{3.0cm}|}
\hline
\textbf{LLM model} &
\textbf{Core model/training} &
\textbf{Network input} &
\textbf{Main difference respect to baseline} &
\textbf{MSE computation} \\
\hline
Baseline &
trainAutoencoder, hidden=64, logsig/logsig &
1024$\times$N vectors &
Minimal sparsity (0.005), L2=0.001 &
mean((recon-x).\textasciicircum{}2) $\rightarrow$ per-sample \\
\hline
SuperGrok &
CAE via trainNetwork (conv/pool + tconv) &
32$\times$32$\times$1$\times$N &
Spatial convolutions, ReLU stack &
mean over [H,W,C] $\rightarrow$ per-image \\
\hline
GPT5.2 &
trainAutoencoder tuned &
1024$\times$N &
Smaller bottleneck + strong sparsity &
same as baseline \\
\hline
GLM5 &
Deeper CAE (trainNetwork) with BN/dropout/LR schedule &
32$\times$32$\times$1$\times$N &
Higher-capacity CAE + explicit regularization (BN/dropout) &
per-sample after reshape \\
\hline
Gemini 3.1 Pro &
trainAutoencoder with noise + strong sparsity &
1024$\times$N &
Noise injection + strong sparsity &
same as baseline \\
\hline
Claude Sonnet 4.6 &
CAE as dlnetwork + custom loop &
32$\times$32$\times$1$\times$N &
Reconstruction + latent centre loss, grad clipping &
per-sample over [H,W,C] \\
\hline
\end{tabularx}
\label{tab:llm_comparison}
\end{table*}
In this setting, the reconstruction \ac{MSE}, computed per sample by averaging over the 1024 pixels, serves directly as the reference metric to check the identity of the transmitter; therefore, any architectural or training change that alters the reconstruction on in-distribution versus out-of-distribution samples is expected to translate into measurable shifts in the \ac{MSE} separation. The LLM-based \ac{AE} designs explore three principal directions. First, {\em SuperGrok} and {\em GLM5} replace the vector autoencoder with convolutional autoencoders trained via \verb|trainNetwork|, operating on 4-D tensors $(32 \times 32 \times 1 \times N)$ and using \verb|ReLU|-based feature extraction. This approach introduces an explicit spatial inductive bias: local patterns in the IQ-histogram plane can be encoded in a compositional way rather than through dense global connections. {\em GLM5} further increases capacity and stabilizes optimization via batch normalization, dropout, and a scheduled learning rate decay, whereas {\em SuperGrok} adopts a compact, symmetric \ac{CAE}. Second, {\em GPT5.2} and {\em Gemini 3.1} retain the \verb|trainAutoencoder| interface but reshape the network design via stronger constraints: {\em GPT5.2} enforces a smaller bottleneck (32 vs. 64) with higher sparsity and controlled scaling, while {\em Gemini 3.1} combines sparsity with denoising (Gaussian perturbations during training only) to improve robustness. Third, {\em Claude Sonnet 4.6} moves to a \verb|dlnetwork| \ac{CAE} trained with a custom Adam loop, by adding a latent center-loss term to the reconstruction loss, thus penalizing how far each sample’s latent vector deviates from the batch mean, while using gradient clipping, thereby encouraging tightly clustered latent representations for in-distribution training data. The convolutional structure and deeper capacity may reduce in-distribution \ac{MSE}, at the expense of general reconstruction; conversely, stronger sparsity/denoising and latent structure should increase discrimination by degrading out-of-distribution reconstructions relative to the baseline.

\section{Performance evaluation}
\label{sec:performance_evaluation}
\begin{figure}
 \centering
 \includegraphics[width=\columnwidth, angle = 0,trim = 0mm 0mm 0mm 0mm]{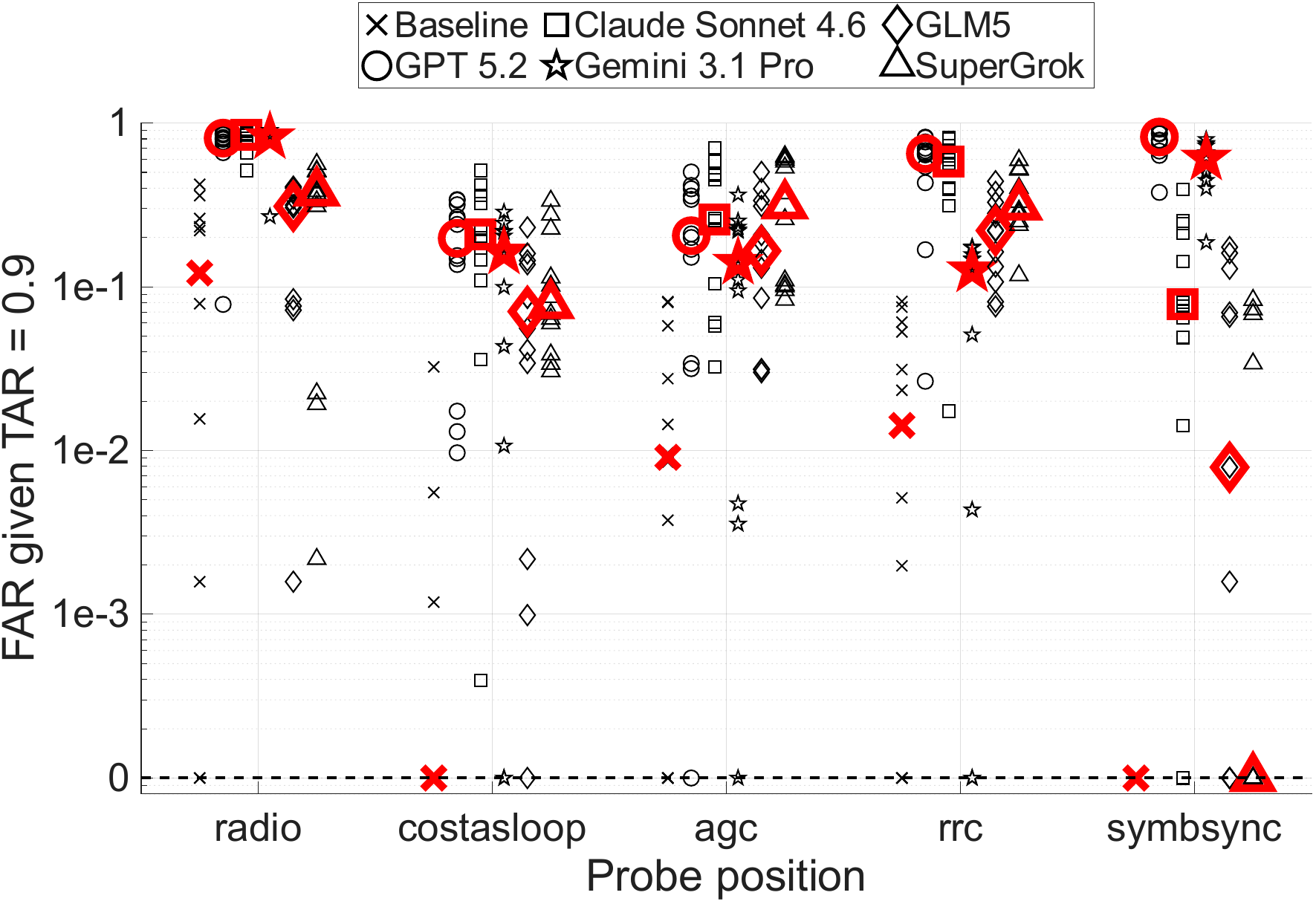}
 \caption{Performance (FAR given TAR = 0.9) as a function of the \ac{RFFI} network and the considered probe. Each black marker corresponds to a different transmitter (out of 12), while the bold red marker corresponds to the median value.}
 \label{fig:all_vs_all}
\end{figure}

Figure~\ref{fig:all_vs_all} summarizes \ac{RFFI} performance by reporting, for each pair (LLM-design, probe), the $\textrm{FAR}$ required to achieve $\textrm{TAR} = 0.9$ across all 12 transmitters (black markers), with the median transmitter highlighted in red. Lower values of $\textrm{FAR}$ indicate better separation between \ac{ID} and \ac{OOD} reconstructions, and $\textrm{FAR} = 0$ corresponds to complete separation. The results show that reliable image-based \ac{RFFI} is highly probe-dependent: only \verb|costasloop| and \verb|symbsync| allow low-FAR operation, whereas \verb|radio|, \verb|agc|, and \verb|rrc| generally require $\textrm{FAR} > 0.1$, indicating substantial overlap between \ac{ID} and \ac{OOD} reconstruction-error distributions. 
A key takeaway from Fig.~\ref{fig:all_vs_all} is that the receiver blocks that are beneficial for communications (carrier recovery and timing alignment) can also be beneficial for \ac{RFF} in this setting: \verb|symbsync| and, to a lesser extent, \verb|costasloop| yield a signal representation where device-specific residuals are more easily captured by reconstruction-based scoring, while \verb|agc| and \verb|rrc| filtering tend to reduce or homogenize amplitude/transient peculiarities, and thus, affecting open-set separability. This behavior is consistent with the progressive ``regularization'' of the \ac{IQ} structure along the chain observed in Sect.~\ref{sec:transmitter_identification}.

The observed ``probe ranking'' suggests a tradeoff between signal variance suppression and fingerprint preservation. Receiver processing can help \ac{RFFI} by removing variations that are not stable transmitter properties (so, not consistent with the radio fingerprint). For example, carrier recovery reduces phase and frequency offsets, and timing recovery aligns samples near symbol centers; both effects can make samples from the same enrolled transmitter more consistent, thereby tightening the \ac{ID} reconstruction-error distribution. At the same time, some processing stages may also suppress residual hardware-dependent characteristics that are useful for distinguishing transmitters. For example, the \verb|agc| block can reduce amplitude-related differences, while \verb|rrc| filtering can smooth high-frequency or transient components that may carry device-specific information. Therefore, a probe is beneficial when it reduces within-transmitter variability more than it removes between-transmitter differences. In our measurements, the probes \verb|costasloop| and especially \verb|symbsync| appear to achieve this empirical balance better than \verb|radio|, \verb|agc|, or \verb|rrc|.

\begin{figure}
 \centering
 \includegraphics[width=0.95\columnwidth, angle = 0,trim = 0mm 0mm 0mm 0mm]{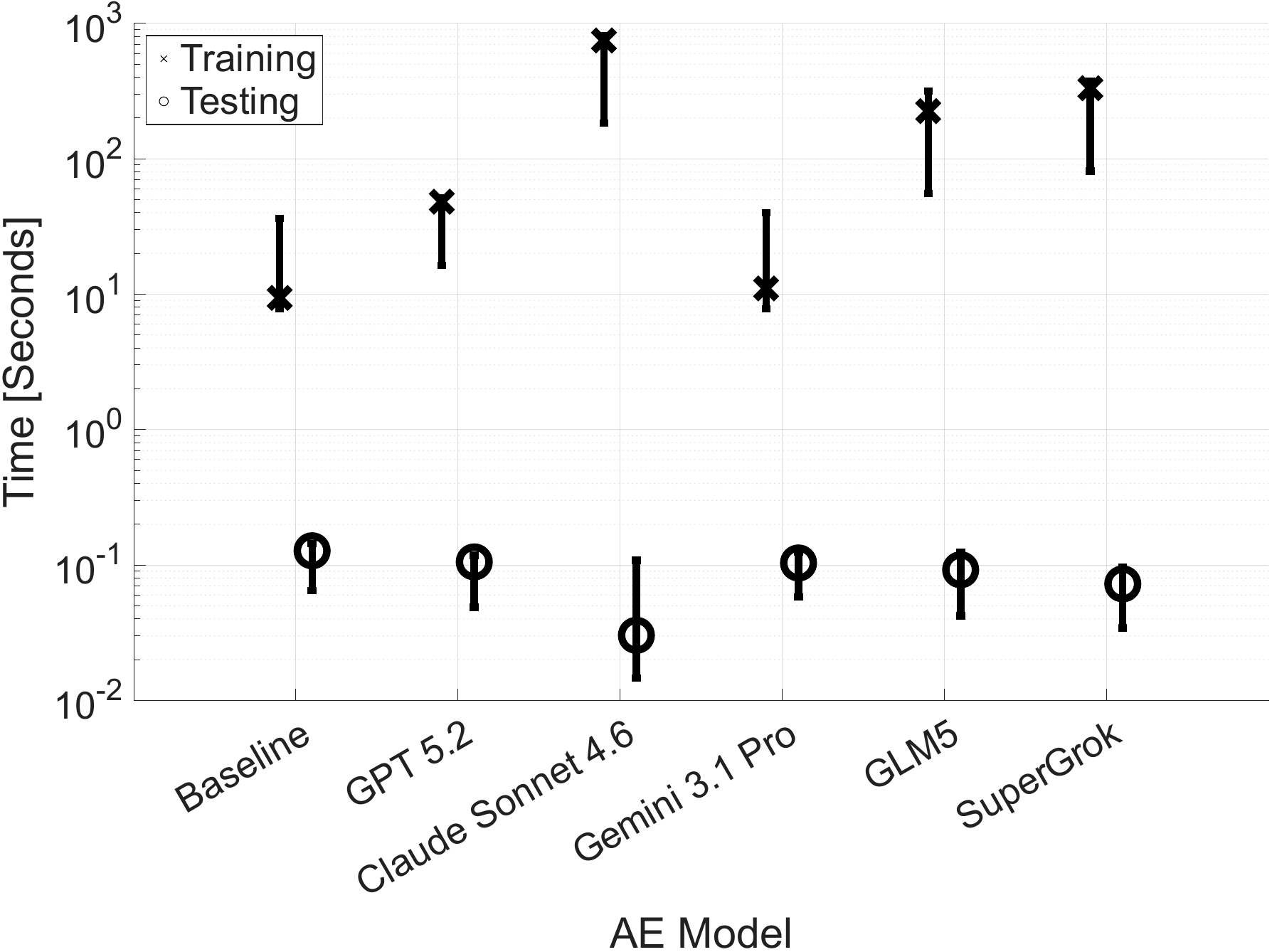}
 \caption{Training and testing time (seconds) as a function of the LLM-based design of the neural network.}
 \label{fig:train_test_time}
\end{figure}

\begin{finding}
    Receiver blocks that reduce impairments for demodulation, e.g., carrier phase/timing uncertainty, can also reduce within-transmitter variability, which in turn can improve \ac{ID}–\ac{OOD} separability under reconstruction-error scoring; however, other blocks, e.g., \verb|agc|/\verb|rrc|, may also remove device-specific features, and thus, affect \ac{RFF}.
\end{finding}
Finally, although LLM-designed networks increase architectural diversity, Fig.~\ref{fig:all_vs_all} indicates that most LLM-driven designs do not improve the performance of the baseline under the same preprocessing and reconstruction-MSE decision rule. A plausible explanation is that higher-capacity or better-regularized models may improve reconstruction not only for \ac{ID} data but also (to some extent) for \ac{OOD} data, thereby affecting to a limited extent the \ac{ID}–\ac{OOD} score gap. This suggests that increasing model capacity or adding stronger regularization can make reconstructions more accurate for both enrolled (\ac{ID}) and non-enrolled (\ac{OOD}) signals, shrinking the separation that open-set detection relies on. SuperGrok at the \verb|symbsync| probe stands out as the only positive case, indicating that a compact convolutional \ac{AE} can better match the 2D histogram structure without improving (too much) \ac{OOD} reconstructions.
\begin{finding}
    Considering the same front-end (IQ-samples $\rightarrow$ images) pipeline and the reconstruction-error score held fixed, \ac{LLM}-based designs do not improve the baseline at the chosen operating point ($\textrm{TAR} = 0.9$) and often require a higher $\textrm{FAR}$---the only exception being SuperGrok with processing samples taken from \verb|symbsync| probe.
\end{finding}
We also analyze training and inference times. Figure~\ref{fig:train_test_time} shows the time to train and test the same amount of data (70\% of the available dataset for each configuration). Indeed, we recall that the train and test sets do not have the same size (70/30), so we scaled the actual test set timings by a factor of $\frac{0.7}{0.3}$. Error-bars show the quantiles 5, 50, and 95 associated with the timings of the training and testing procedures for the baseline design and the different \ac{LLM}-suggested neural networks. While the differences in inference times are not significant, training is more significantly affected by the neural network design. Indeed, the training time spans between 10~s (median) for the baseline to about 748~s (median) for Claude Sonnet 4.6.

\section{Probe-Aware Enrollment and Cross-Probe Transferability}
\label{sec:probes}
{\bf Multi-probe enrollment.} The methodology discussed so far focused on the enrollment (single-TX, probe). However, the information coming from several probes could potentially enhance \ac{RFFI} performance. Under this assumption, in this section, we examine all probe positions during enrollment. Concretely, for a given enrolled transmitter, i.e., $\bar{T} \in \{1, \ldots, 12\}$, we train each model (baseline + all LLM-designed \acp{AE}) using the union of the available data coming from all probes, i.e., $P \in \{\verb|radio|, \verb|costasloop|, \verb|agc|, \verb|rrc|, \verb|symbsync|\}$. Testing is then performed on the remainder of the transmitters, i.e., $T \in \{1, \ldots, 12\} \setminus \bar{T}$, again considering all probes altogether, using the same reconstruction‑error scoring rule.

\begin{figure}
 \centering
 \includegraphics[width=0.95\columnwidth, angle = 0,trim = 0mm 0mm 0mm 0mm]{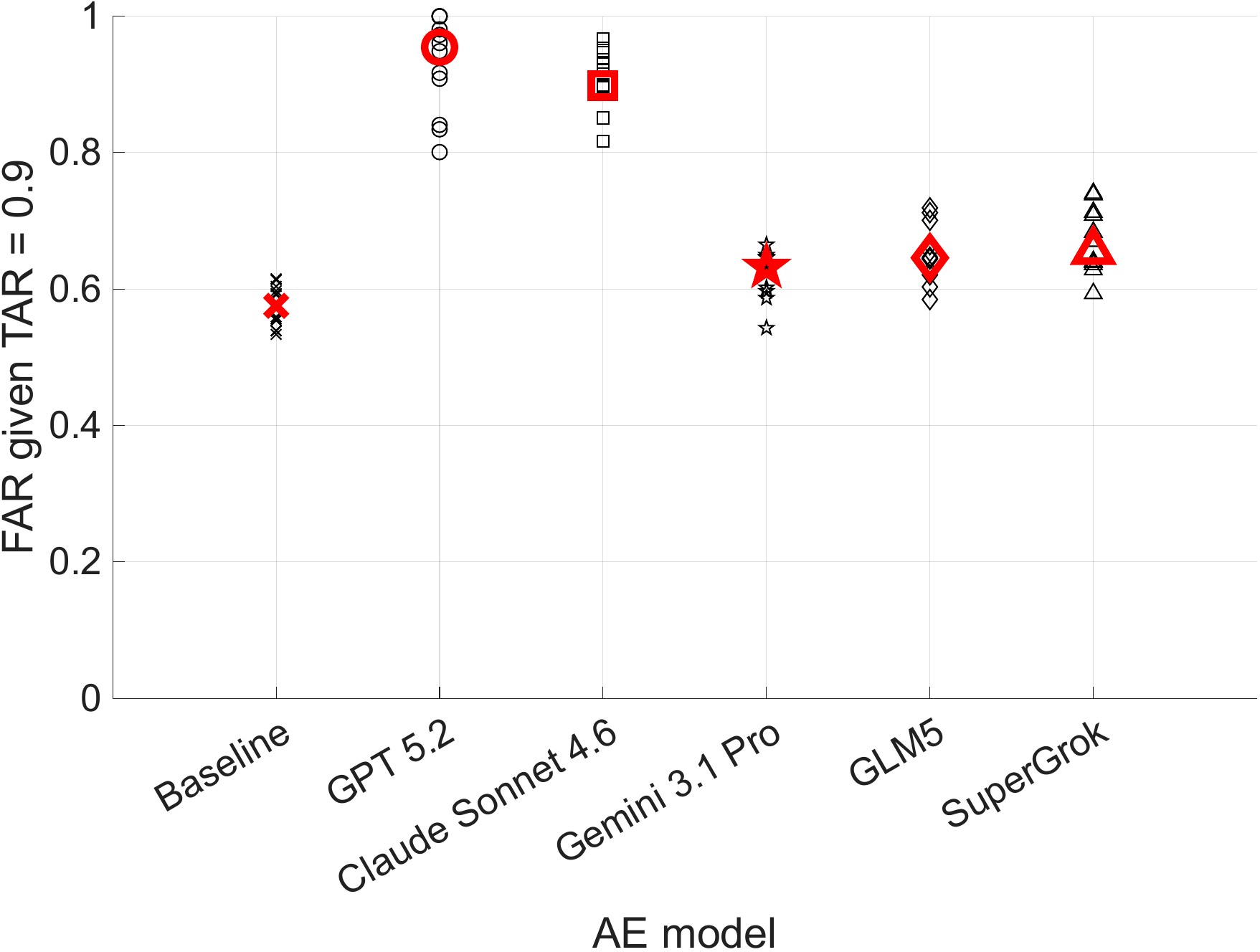}
 \caption{Performance (FAR given TAR = 0.9) as a function of the \ac{AE} network, considering a model trained on all the probes. Each black marker corresponds to a different transmitter (out of 12), while the red marker highlights the median value.}
 \label{fig:modellone_all_vs_all}
\end{figure}

Figure~\ref{fig:modellone_all_vs_all} shows the $\textrm{FAR}$ at $\textrm{TAR} = 0.9$ as a function of the considered LLM-based model. The black markers represent the operating point ($\textrm{FAR}$ at $\textrm{TAR} = 0.9$) for each transmitter, and we highlight the median in red.
The outcome is consistently poor: \ac{ID} and \ac{OOD} reconstruction‑error distributions heavily overlap, producing \ac{ROC} curves close to a random-guess behavior and operating points that require large $\textrm{FAR}$ to sustain a reasonable $\textrm{TAR} = 0.9$. This happens across the baseline and all LLM-generated architectures, indicating that the issue primarily stems from the source mismatch introduced by mixing probes rather than from the specific \ac{AE} design.

This behavior can be explained by noting that each probe corresponds to a distinct signal generated by a distinct receiver transformation. Combining the probes makes the \ac{ID} samples (more) multimodal and dominated by probe-specific structure; i.e., the variability introduced by the processing stages is larger than the device-specific features we want the model to use as a fingerprint. 
Under a reconstruction objective, i.e., minimization of the \ac{MSE}, the autoencoder is biased to learn a representation that is probe-robust and that captures shared features; as a side effect, it tends to reconstruct \ac{OOD} samples better than desired (shrinking the \ac{ID}–\ac{OOD} gap), while also increasing the \ac{ID} errors due to cross-probe heterogeneity.

Overall, our findings reinforce a practical conclusion: multi‑probe pooling during enrollment is counterproductive for reconstruction‑error open‑set \ac{RFF}: probe-aware strategies, e.g., per‑probe enrollment, achieve better performance.

\begin{figure}
 \centering
 \includegraphics[width=0.95\columnwidth, angle = 0,trim = 0mm 0mm 0mm 0mm]{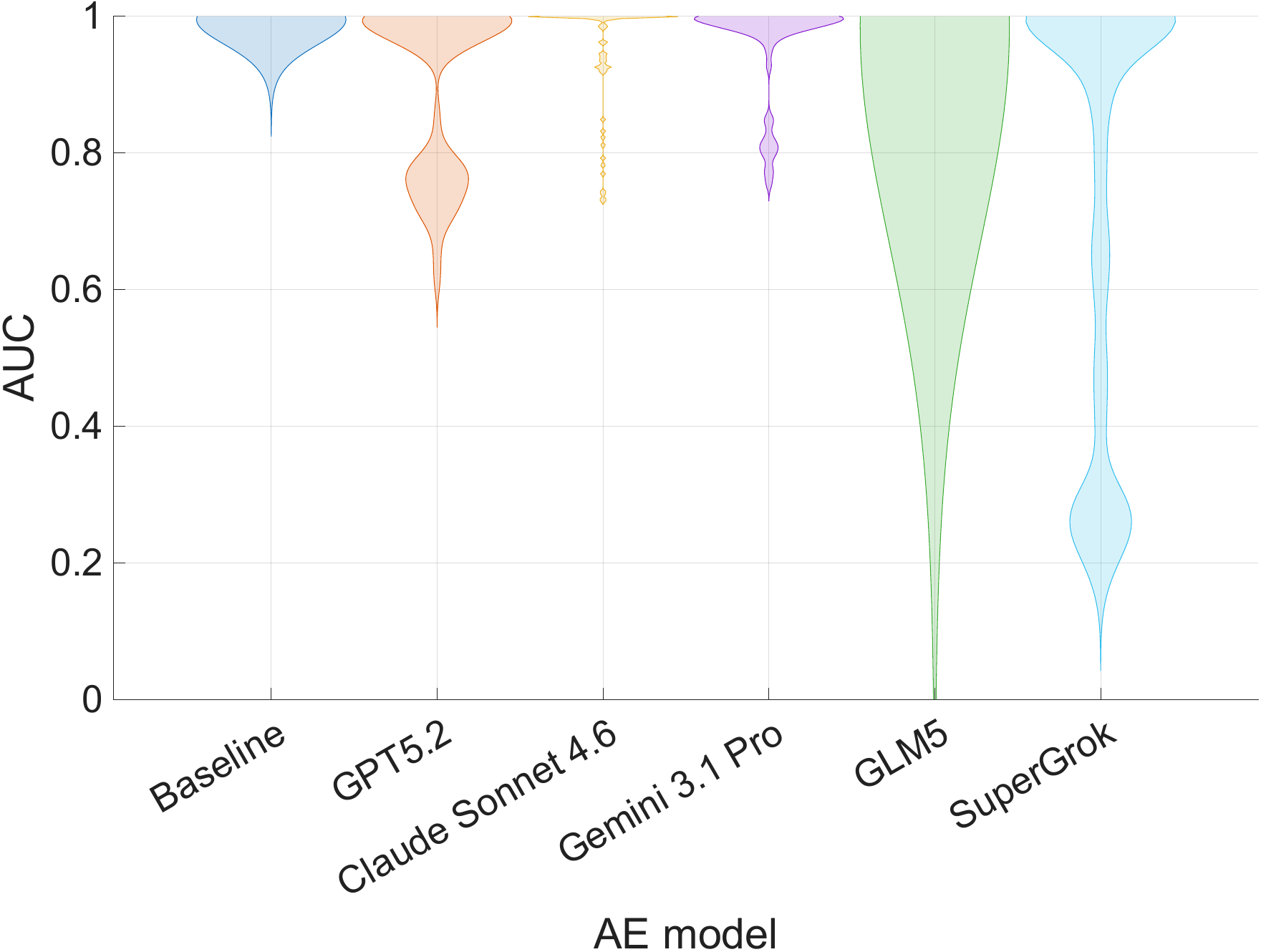}
 \caption{\ac{AUC} training on data coming from a specific probe and testing on any other probe, considering the same transmitter as a function of the LLM model.}
 \label{fig:probe}
\end{figure}

{\bf Cross-Probe Transferability.} Figure~\ref{fig:probe} reports a cross-probe analysis designed to quantify to what extent the receiver-chain probe reshapes the learned representation for a given transmitter. 
For each transmitter and enrollment probe, we train an \ac{AE} using only data from the pair. We then evaluate the same trained model on: (i) (disjoint) data from the same pair transmitter-probe, treated as \ac{ID}, and (ii) data from the same transmitter but different probes, treated as \ac{OOD}. This procedure is repeated for all transmitter/probe combinations, and the resulting \ac{ROC} \ac{AUC} summarizes how separable the ID vs. cross-probe \ac{OOD} reconstruction-error distributions are for each model family. An \ac{AUC} close to 1 indicates that the reconstruction-error score reliably separates samples collected at different probes, even when the transmitter is unchanged. Thus, this entails that the autoencoder trained on a given probe systematically assigns higher reconstruction errors to samples from any other probe. In turn, this result implies that probe-dependent transformations dominate the features seen by the reconstruction model, consistent with the idea that each receiver block reshapes (or partially suppresses) device-dependent structures. The {\em baseline} is characterized by \ac{AUC}$\approx 1$ almost everywhere, reinforcing the earlier conclusion that probes reshape the fingerprint representation. The same values (with some anomalies) can be seen in {\em GPT5.2}, {\em Claude Sonnet 4.6}, and {\em Gemini 3.1 Pro}. 
We investigated the cases of {\em GLM5} and {\em SuperGrok} in more detail by manually inspecting the \ac{AUC} values and the pairs of probe-transmitter affected by the anomalies. Our analysis shows that \ac{AUC}$<<1$ is best interpreted as a failure of the reconstruction-error across probes, rather than a metric to prove feature preservation. Indeed, both models, i.e., {\em GLM5} and {\em SuperGrok}, are still characterized by high AUC ($\approx 1$), but with isolated (non-consistent) anomalies, where \ac{OOD} samples are easier to reconstruct than \ac{ID} samples (inversion). Such a behavior (in particular for probes \verb|rrc| and \verb|agc|) is consistent and can be explained by recalling Sect.~\ref{sec:performance_evaluation}, i.e., improving ``general'' reconstruction can also improve \ac{OOD} reconstruction. {\em SuperGrok} is representative, especially considering \verb|rrc| and \verb|agc|, as it tends to homogenize or suppress device-specific artifacts, increasing the overlap between \ac{ID} and \ac{OOD} samples.

\section{Discussion}
\label{sec:discussion}

This work represents a controlled experimental probe-sensitivity study. We isolate an under-reported design variable: the receiver-chain point at which IQ samples are collected. The central claim is that this choice substantially alters reconstruction-based open-set \ac{RFFI} and, therefore, should be explicitly reported, controlled, and evaluated. 
Accordingly, \ac{RFFI} solutions should not treat IQ collection location as an incidental logging detail. A pipeline that appears ineffective at one probe may become viable (or a top performer) at another, while pooling probes can degrade performance. Therefore, \ac{RFFI} works should report probe location alongside model architecture and dataset details.

{\bf The tap matters more than the \ac{AE}.} Our results show that, for reconstruction-error image-based open-set \ac{RFFI}, the data source, i.e., where the signal is observed in the receiver chain, matters more than the configuration of the image-based \ac{AE}. Across five probe positions, we observe that timing recovery (\verb|symbsync|) and, to a lesser extent, carrier recovery (\verb|costasloop|),produce representations that enable low-$\textrm{FAR}$ operation at $\textrm{TAR} = 0.9$, while the other processing stages, i.e., \verb|radio|, \verb|agc|, and \verb|rrc|, tend to exhibit a substantial overlap between \ac{ID} and \ac{OOD} reconstruction-error distributions. Thus, processing blocks that reduce phase and timing uncertainty can also reduce within-transmitter variability, thereby tightening the \ac{ID} error distribution and improving MSE-based discrimination. 

{\bf Not All IQs Are Equal.} A second, closely related implication is that the receiver processing chain is not transparent to \ac{RFFI}. Standard communications-oriented transformations reshape the joint \ac{IQ} distribution in ways that can either amplify or suppress the features that \ac{RFFI} relies on. In our setting, \verb|symbsync| and \verb|costasloop| appear to be the best data sources: they reduce the variation (timing/phase), yet preserve enough consistent device-dependent features for a reconstruction-based model to treat \ac{OOD} samples as ``harder to reconstruct'' than \ac{ID} samples. 

{\bf LLM-based \ac{AE} design.} Within this probe-aware framing, our LLM-generated image-based \ac{AE} variants serve primarily as a controlled stress test of the hypothesis that more complex architectures could align performance across probes and possibly improve discrimination. When the front-end IQ-to-image pipeline and the reconstruction-MSE decision rule are held fixed, different \ac{AE} designs do not improve operating-point performance, and frequently worsen it. {\em SuperGrok} at \verb|symbsync| suggests that some inductive biases---a convolutional structure aligned with the 2D histogram representation---can improve \ac{ID} reconstruction without proportionally improving \ac{OOD} reconstruction, paving the way for future analyses.

{\bf Takeaway.} Our timing results bias the trade-off even further: while inference cost is comparable across designs, the training time increases significantly, without any significant gain in separability (Fig.~\ref{fig:train_test_time}). For an operational \ac{RFFI} pipeline that must be retrained across devices, time, or environments, this finding suggests a straight priority order: (1) choose the probe/representation that maximizes separability, then (2) keep the \ac{AE} design as simple as possible unless there is strong evidence that added complexity improves the \ac{ID}–\ac{OOD} gap for that specific probe.

{\bf Limitations.} 
Our evaluation considers a specific measurement configuration (hardware, receiver chain, modulation, dataset scale) and a fixed representation, i.e., $32 \times 32$ histogram ``RF images'' with a fixed chunk size and binning policy. Although the probe ranking could change considering different modulations, receiver implementations, quantization, over-the-air channels, or front-end impairments, the different shapes characterizing \ac{IQ} samples collected at different probes (see Fig.~\ref{fig:iq_examples}) are very likely to yield the same main finding, i.e., the dependence of \ac{RFFI} performance on the specific probe. 
Moreover, we do not claim that the absolute probe ranking reported in this paper holds unchanged across all modulations, transmitters, receivers, channels, or over-the-air deployments. Future work could investigate how much such ranking holds with more complex high-order modulation schemes.
Furthermore, we report performance primarily via $\textrm{FAR}$ at $\textrm{TAR} = 0.9$. Although this choice does not capture all operating scenarios, it is appropriate for operational comparison at a fixed quality target. 
Finally, we do not consider physical-layer spoofing attacks, where an adversary intentionally biases the transmitted \ac{IQ} samples to resemble the enrolled transmitter at a chosen probe. Such attacks are distinct from the non-enrolled-transmitter \ac{OOD} setting studied here. Our results characterize probe dependence under (non-malicious) transmitter mismatch, rather than robustness against an attacker with knowledge of the receiver chain and the \ac{RFFI} model. Evaluating whether probe choices that improve separability also increase or decrease spoofing resistance is an important direction for future work.

\section{Conclusion}
\label{sec:conclusion}
In this work, we assess how strongly the performance of \acl{PHY}-layer, open-set, image-based \ac{RFFI} depends on the receiver-chain \emph{probe} used for data collection, i.e., the stage at which the \ac{IQ} samples are acquired. By logging IQ data at five tap points along a GNU Radio \ac{BPSK} receiver chain, we showed that \ac{RFFI} performance is strongly probe-dependent. In particular, timing recovery (\verb|symbsync|)---and, to a lesser extent, carrier recovery (\verb|costasloop|)---supports low false acceptance at a target operating point, e.g., $\textrm{FAR}$ at $\textrm{TAR} = 0.9$, while other processing stages often exhibit substantial \ac{ID}-\ac{OOD} score overlap. These results indicate that receiver blocks reducing phase/timing uncertainty can tighten within-transmitter variability and improve separability, whereas other blocks may suppress device-specific features. As a secondary contribution, we benchmarked several LLM-designed image-based \acp{AE} for \ac{RFFI} under a controlled approach that keeps the preprocessing pipeline and MSE scoring rule fixed while changing only the network design. Most LLM-generated architectures did not improve the baseline at the chosen operating point, but just degraded it, with only one notable exception (SuperGrok for the probe \verb|symbsync|). Overall, the key takeaway is practical: ``where you tap'' in the receiver chain can matter more than the \ac{AE} architecture for reconstruction-based open-set \ac{RFFI}.

\bibliographystyle{IEEEtran}
\balance
\bibliography{main}

\end{document}